\documentclass[proceedings]{JHEP37}
\usepackage{eurosym}
\usepackage{amssymb}
\usepackage{amsfonts}
\usepackage{amsmath,epsfig}
\usepackage{graphicx}

\newbox\mybox

\newcommand\fverb{\setbox\mybox=\hbox\bgroup\verb}
\newcommand\fverbdo{\egroup\medskip\noindent\fbox{\unhbox\mybox}\ }
\newcommand\fverbit{\egroup\item[\fbox{\unhbox\mybox}]}
\conference{Multicomplex solitons}
\abstract{We discuss integrable extensions of real nonlinear wave equations with multi-soliton solutions, to their bicomplex, 
quaternionic, coquaternionic and octonionic versions. In particular, we investigate these variants for the local and nonlocal 
Korteweg-de Vries equation and elaborate on how multi-soliton solutions with various types of novel qualitative behaviour can be constructed. 
Corresponding to the different multicomplex units in these extensions, real, hyperbolic or imaginary, the wave equations and their solutions exhibit 
multiple versions of antilinear or $\mathcal{PT}$-symmetries. Utilizing these symmetries forces certain components of the conserved quantities to vanish, so that one may 
enforce them to be real. We find that symmetrizing the noncommutative equations is equivalent to imposing a $\mathcal{PT}$-symmetry for a newly defined imaginary unit 
from combinations of imaginary and hyperbolic units in the canonical representation.}

\title{Multicomplex solitons}
\author{Julia Cen and Andreas Fring \\
Department of Mathematics, City, University of London,\\
Northampton Square, London EC1V 0HB, UK \\
E-mail: julia.cen.1@city.ac.uk, a.fring@city.ac.uk}

\input{tcilatex}
\begin{document}

\section{Introduction}

The underlying mathematical structure of quantum mechanics, a Hilbert space
over the field of complex numbers, can be generalized and modified in
various different ways. One may for instance re-define the inner product of
the Hilbert space or alter, typically enlarge, the field over which this
space is defined. The first approach has been pursued successfully since
around twenty years \cite{Bender:1998ke}, when it was first realized that
the modification of the inner product allows to include non-Hermitian
Hamiltonians into the framework of a quantum mechanical theory. When these
non-Hermitian Hamiltonians are $\mathcal{PT}$-symmetric/quasi-Hermitian \cite%
{Urubu,Benderrev,Alirev} they possess real eigenvalues when their
eigenfunctions are also $\mathcal{PT}$-symmetric or pairs of complex
conjugate eigenvalues when the latter is not the case. The reality of the
spectrum might only hold in some domain of the coupling constant, but break
down at what is usually referred to as an exceptional point when at least
two eigenvalues coalesce. Higher order exceptional points may occur for
larger degeneracies. In order to unravel the structure of the neighbourhood
of these points one can make use of the second possibility of
generalizations of standard quantum mechanics and change the type of fields
over which the Hilbert space is defined. This view helps to understand the
bifurcation structure at these points and has been recently investigated for
the analytically continued Gross-Pitaevskii equation with bicomplex
interaction terms \cite{BiGP1,BiGP2,BiGP3}. In a similar spirit, systems
with finite dimensional Hilbert spaces have been formulated over Galois
fields \cite{vourdas}. Hyperbolic extensions of the complex Hilbert space
have been studied in \cite{xuegang2000hyperbolic}. The standard Schr\"{o}%
dinger equation was bicomplexified in \cite{bagchibicomplex} and further
studied in \cite{BiHam1,BiHam2,BiHam3,theaker2017mult}. Quaternionic and
coquaternionic quantum mechanics and quantum field theory have been studied
for a long time, see e.g. \cite%
{finkelstein1962foundations,girard1984quaternion,adler1995quaternionic},
mainly motivated by the fact that they may be related to various groups and
algebras that play a central role in physics, such as $SO(3)$, the Lorentz
group, the Clifford algebra or the conformal group. Recently it was
suggested that they \cite{brodycoquaternions} provide a unifying framework
for complexified classical and quantum mechanics. Octonionic Hilbert spaces
have been utilized for instance in the study of quark structures \cite%
{gunaydin1973quark}.

Drawing on various relations between the quantum mechanical setting and
classical integrable nonlinear systems that possess soliton solutions, such
as the formal identification of the $L$ operator in a Lax pair as a
Hamiltonian, many of the above possibilities can also been explored in the
latter context. Most direct are the analogues of the field extensions.
Previously we demonstrated \cite{CenFring,cen2016time} that one may
consistently extend real classical integrable nonlinear systems to the
complex domain by maintaining the reality of the energy. Here we go further
and investigate multicomplex versions of these type of nonlinear equations.
We demonstrate how these equations can be solved in several multicomplex
settings and study some of the properties of the solutions. We explore three
different possibilities to construct solutions that are not available in a
real setting, i) using multicomplex shifts in a real solutions, ii)
exploiting the complex representations by defining a new imaginary unit in
terms of multicomplex ones and iii) exploiting the idempotent
representation. We take $\mathcal{PT}$-symmetry as a guiding principle to
select out physically meaningful solutions with real conserved quantities,
notably real energies. We clarify the roles played by the different types of 
$\mathcal{PT}$-symmetries. For the noncommutative versions, that is
quaternionic, coquaternionic and octonionic, we find that imposing certain $%
\mathcal{PT}$-symmetries corresponds to symmetrizing the noncommutative
terms in the nonlinear differential equations.

Our manuscript is organized as follows: In section 2 we discuss the
construction of bicomplex multi-solitons for the standard Korteweg de-Vries
(KdV) equation and its nonlocal variant. We present two different types of
construction schemes leading to solutions with different types of $\mathcal{%
PT}$-symmetries. We demonstrate that the conserved quantities constructed
from these solutions, in particular the energy, are real. In section 3, 4
and 5 we discuss solution procedures for noncommutative versions of the KdV
equation in quaternionic, coquaternionic and octonionic form, respectively.
Our conclusions are stated in section 6.

\section{Bicomplex solitons}

\subsection{Bicomplex numbers and functions}

We start by briefly recalling some key properties of bicomplex numbers and
functions to settle our notations and conventions. Denoting the field of
complex numbers with imaginary unit $\imath $ as 
\begin{equation}
\mathbb{C}(\imath )=\left\{ x+\imath y~|~x,y\in \mathbb{R}\right\} ,
\end{equation}%
the \emph{bicomplex numbers} $\mathbb{B}$ form an algebra over the complex
numbers admitting various equivalent types of representations 
\begin{eqnarray}
\mathbb{B} &\mathbb{=}&\left\{ z_{1}+\jmath z_{2}~|~z_{1},z_{2}\in \mathbb{C}%
(\imath )\right\} ,  \label{B1} \\
&=&\left\{ w_{1}+\imath w_{2}~|~w_{1},w_{2}\in \mathbb{C}(\jmath )\right\} ,
\label{B2} \\
&=&\left\{ a_{1}\ell +a_{2}\imath +a_{3}\jmath
+a_{4}k~|~a_{1},a_{2},a_{3},a_{4}\in \mathbb{R}\right\} ,  \label{B3} \\
&=&\left\{ v_{1}e_{1}+v_{2}e_{2}~|~v_{1}\in \mathbb{C}(\imath ),v_{2}\in 
\mathbb{C}(\jmath )\right\} .  \label{B4}
\end{eqnarray}%
The \emph{canonical basis} is spanned by the units $\ell $, $\imath $, $%
\jmath $, $k$, involving the two \emph{imaginary units} $\imath $ and $%
\jmath $ with $\imath ^{2}=\jmath ^{2}=-1$, so that the representations in
equations (\ref{B1}) and (\ref{B2}) naturally prompt the notion to view
these numbers as a doubling of the complex numbers. The real unit $\ell $
and the \emph{hyperbolic unit} $k=\imath \jmath $ square to $1$, $\ell
^{2}=k^{2}=1$. The multiplication of these units is commutative with further
products in the Cayley multiplication table being $\ell \imath =\imath $, $%
\ell \jmath =\jmath $, $\ell k=k$, $\imath k=-\jmath $, $\jmath k=-\imath $.
The \emph{idempotent representation} (\ref{B4}) is an orthogonal
decomposition obtained by using the orthogonal idempotents%
\begin{equation}
e_{1}:=\frac{1+k}{2},\qquad \text{and\qquad }e_{2}:=\frac{1-k}{2},
\end{equation}%
with properties $e_{1}^{2}=e_{1}$, $e_{2}^{2}=e_{2}$, $e_{1}e_{2}=0$ and $%
e_{1}+e_{2}=1$. All four representations (\ref{B1}) - (\ref{B4}) are
uniquely related to each other. For instance, given a bicomplex number in
the canonical representation (\ref{B3}) in the form%
\begin{equation}
n_{a}=a_{1}\ell +a_{2}\imath +a_{3}\jmath +a_{4}k,  \label{ar}
\end{equation}%
the equivalent representations (\ref{B1}), (\ref{B3}) and (\ref{B4}) are
obtained with the identifications%
\begin{equation}
\begin{array}{ll}
z_{1}=a_{1}+\imath a_{2}, & z_{2}=a_{3}+\imath a_{4}, \\ 
w_{1}=a_{1}+\jmath a_{3}, & w_{2}=a_{2}+\jmath a_{4}, \\ 
v_{1}^{a}=(a_{1}+a_{4})\ell +(a_{2}-a_{3})\imath ~,~~~~~\  & 
v_{2}^{a}=(a_{1}-a_{4})\ell +(a_{2}+a_{3})\jmath .%
\end{array}
\label{rel}
\end{equation}%
Arithmetic operations are most elegantly and efficiently carried out in the
idempotent representation (\ref{B4}). For the composition of two arbitrary
numbers $n_{a}$ and $n_{b}$ we have 
\begin{equation}
n_{a}\circ n_{b}=v_{1}^{a}\circ v_{1}^{b}e_{1}+v_{2}^{a}\circ
v_{2}^{b}e_{2}\quad ~~~\text{with }\circ \equiv \pm ,\cdot ,\div .
\label{comp}
\end{equation}%
The \emph{hyperbolic numbers }(or\emph{\ split-complex numbers}) $\mathbb{D=}%
\left\{ a_{1}\ell +a_{4}k~|~a_{1},a_{4}\in \mathbb{R}\right\} $ are an
important special case of $\mathbb{B}$ obtained in the absence of the
imaginary units $\imath $ and $\jmath $, or when taking $a_{2}=a_{3}=0$.

The same arithmetic rules as in (\ref{comp}) then apply to \emph{bicomplex
functions}. In what follows we are most interested in functions depending on
two real variables $x$ and $t$ of the form $f(x,t)=\ell p(x,t)+\imath
q(x,t)+\jmath r(x,t)+ks(x,t)\in \mathbb{B}$ involving four real fields $%
p(x,t)$, $q(x,t)$, $r(x,t)$, $s(x,t)\in \mathbb{R}$. Having kept the
functional variables real, we also keep our differential real, so that we
can differentiate $f(x,t)$ componentwise as $\partial _{x}f(x,t)=\ell
\partial _{x}p(x,t)+\imath \partial _{x}q(x,t)+\jmath \partial
_{x}r(x,t)+k\partial _{x}s(x,t)$ and similarly for $\partial _{t}f(x,t)$.
For further properties of bicomplex numbers and functions, such as for
instance computing norms, see for instance \cite%
{price1991,davenport1996commutative,rochon,luna2012bicomplex}.

\subsection{$\mathcal{PT}$-symmetric bicomplex functions and conserved
quantities}

As there are two different imaginary units, there are three different types
of conjugations for bicomplex numbers, corresponding to conjugating only $%
\imath $, only $\jmath $ or conjugating both $\imath $ and $\jmath $
simultaneously. This is reflected in different symmetries that leave the
Cayley multiplication table invariant. As a consequence we also have three
different types of bicomplex $\mathcal{PT}$-symmetries, acting as%
\begin{eqnarray}
\mathcal{PT}_{\imath \jmath } &:&\ell \rightarrow \ell ,\imath \rightarrow
-\imath ,\jmath \rightarrow -\jmath ,k\rightarrow k,~x\rightarrow
-x,t\rightarrow -t,\ \   \label{PT1} \\
\mathcal{PT}_{\imath k} &:&\ell \rightarrow \ell ,\imath \rightarrow -\imath
,\jmath \rightarrow \jmath ,k\rightarrow -k,~x\rightarrow -x,t\rightarrow -t,
\label{PT2} \\
\mathcal{PT}_{\jmath k} &:&\ell \rightarrow \ell ,\imath \rightarrow \imath
,\jmath \rightarrow -\jmath ,k\rightarrow -k,~x\rightarrow -x,t\rightarrow
-t,\   \label{PT3}
\end{eqnarray}%
see also \cite{bagchibicomplex}. When decomposing the bicomplex energy
eigenvalue of a bicomplex Hamiltonian $H$ in the time-independent Schr\"{o}%
dinger equation, $H\psi =E\psi $, as $E=E_{1}\ell +E_{2}\imath +E_{3}\jmath
+E_{4}k$, Bagchi and Banerjee argued in \cite{bagchibicomplex} that a $%
\mathcal{PT}_{\imath k}$-symmetry ensures that $E_{2}=E_{4}=0$, a $\mathcal{%
PT}_{\jmath k}$-symmetry forces $E_{3}=E_{4}=0$ and a $\mathcal{PT}_{\imath
\jmath }$-symmetry sets $E_{2}=E_{3}=0$. In \cite%
{CenFring,cen2016time,cen2018asymptotic} we argued that for complex soliton
solutions the $\mathcal{PT}$-symmetries together with the integrability of
the model guarantees the reality of all physical conserved quantities. One
of the main concerns in this section is to investigate the roles played by
the symmetries\ (\ref{PT1})-(\ref{PT3}) for the bicomplex soliton solutions
and to clarify whether the implications are similar as observed in the
quantum case.

Decomposing a density function for any conserved quantity as 
\begin{equation}
\rho (x,t)=\ell \rho _{1}(x,t)+\imath \rho _{2}(x,t)+\jmath \rho
_{3}(x,t)+k\rho _{4}(x,t)\in \mathbb{B},
\end{equation}%
and demanding it to be $\mathcal{PT}$-invariant, it is easily verified that
a $\mathcal{PT}_{\imath k}$-symmetry implies that $\rho _{1}$, $\rho _{3}$
and $\rho _{2}$, $\rho _{4}$ are even and odd functions of $x$,
respectively. A $\mathcal{PT}_{\jmath k}$-symmetry forces $\rho _{1}$, $\rho
_{2}$ and $\rho _{3}$, $\rho _{4}$ to even and odd in $x$, respectively and
a $\mathcal{PT}_{\imath \jmath }$-symmetry makes $\rho _{1}$, $\rho _{4}$
and $\rho _{2}$, $\rho _{3}$ even and odd in $x$, respectively. The
corresponding conserved quantities must therefore be of the form%
\begin{equation}
Q=\dint\nolimits_{-\infty }^{\infty }\rho (x,t)dx=\left\{ 
\begin{array}{ll}
Q_{1}\ell +Q_{3}\jmath & \text{for }\mathcal{PT}_{\imath k}\text{-symmetric }%
\rho \\ 
Q_{1}\ell +Q_{2}\imath \qquad & \text{for }\mathcal{PT}_{\jmath k}\text{%
-symmetric }\rho \\ 
Q_{1}\ell +Q_{4}k & \text{for }\mathcal{PT}_{\imath \jmath }\text{-symmetric 
}\rho%
\end{array}%
\right. ,
\end{equation}%
where we denote $Q_{i}:=\dint\nolimits_{-\infty }^{\infty }\rho _{i}(x,t)dx$
with $i=1,2,3,4$. Thus we expect the same property that forces certain
quantum mechanical energies to vanish to hold similarly for all classical
conserved quantities. We only regard $Q_{1}$ and $Q_{4}$ as physical, so
that only a $\mathcal{PT}_{\imath \jmath }$-symmetric system is guaranteed
to be physical.

\subsection{The bicomplex Korteweg-de Vries equation}

Using the multiplication law (\ref{comp}) for bicomplex functions, the KdV
equation for a bicomplex field in the canonical form 
\begin{equation}
u(x,t)=\ell p(x,t)+\imath q(x,t)+\jmath r(x,t)+ks(x,t)\in \mathbb{B},
\label{ub}
\end{equation}%
can either be viewed as a set of coupled equations for the four real fields $%
p(x,t)$, $q(x,t)$, $r(x,t)$, $s(x,t)\in \mathbb{R}$ 
\begin{equation}
u_{t}+6uu_{x}+u_{xxx}=0\quad \Leftrightarrow \quad \left\{ 
\begin{array}{r}
p_{t}+6pp_{x}-6qq_{x}-6rr_{x}+6ss_{x}+p_{xxx}=0 \\ 
q_{t}+6qp_{x}+6pq_{x}-6sr_{x}-6rs_{x}+q_{xxx}=0 \\ 
r_{t}+6rp_{x}+6pr_{x}-6qs_{x}-6sq_{x}+r_{xxx}=0 \\ 
s_{t}+6sp_{x}+6ps_{x}+6qr_{x}+6rq_{x}+s_{xxx}=0%
\end{array}%
\right. ,  \label{KdV}
\end{equation}%
or when using the representation (\ref{B4}) as a couple of complex KdV
equations%
\begin{equation}
v_{t}+6vv_{x}+v_{xxx}=0,\qquad \text{and\qquad }w_{t}+6ww_{x}+w_{xxx}=0,
\label{SKdV}
\end{equation}%
related to the canonical representation as%
\begin{eqnarray}
v(x,t) &=&\left[ p(x,t)+s(x,t)\right] +\imath \left[ q(x,t)-r(x,t)\right]
\in \mathbb{C}(\imath ), \\
w(x,t) &=&\left[ p(x,t)-s(x,t)\right] +\jmath \left[ q(x,t)+r(x,t)\right]
\in \mathbb{C}(\jmath ).
\end{eqnarray}%
We recall that we keep here our space and time variables, $x$ and $t$, to be
both real so that also the corresponding derivatives $\partial _{x}$ and $%
\partial _{t}$ are not bicomplexified.

When acting on the component functions the $\mathcal{PT}$-symmetries (\ref%
{PT1})-(\ref{PT3}) are implemented in (\ref{KdV}) as 
\begin{eqnarray}
\mathcal{PT}_{\imath \jmath } &:&x\rightarrow -x,\ t\rightarrow
-t,p\rightarrow p,q\rightarrow -q,r\rightarrow -r,s\rightarrow
s,u\rightarrow u,~\ \  \\
\mathcal{PT}_{\imath k} &:&x\rightarrow -x,\ t\rightarrow -t,p\rightarrow
p,q\rightarrow -q,r\rightarrow r,s\rightarrow -s,u\rightarrow u, \\
\mathcal{PT}_{\jmath k} &:&x\rightarrow -x,\ t\rightarrow -t,p\rightarrow
p,q\rightarrow q,r\rightarrow -r,s\rightarrow -s,u\rightarrow u,
\end{eqnarray}%
ensuring that the KdV-equation remains invariant for all of the
transformations. Notice that the representation in (\ref{SKdV}) remains only
invariant under $\mathcal{PT}_{\imath \jmath }$, but does not respect the
symmetries $\mathcal{PT}_{\imath k}$ and $\mathcal{PT}_{\jmath k}$.

We observe that (\ref{KdV}) allows for a scaling of space by the hyperbolic
unit $k$ as $x\rightarrow kx$, leading to a new type of KdV-equation with $%
u\rightarrow h$

\begin{equation}
kh_{t}+6hh_{x}+h_{xxx}=0\quad \Leftrightarrow \quad \left\{ 
\begin{array}{r}
s_{t}+6pp_{x}-6qq_{x}-6rr_{x}+6ss_{x}+p_{xxx}=0 \\ 
r_{t}-6qp_{x}-6pq_{x}+6sr_{x}+6rs_{x}-q_{xxx}=0 \\ 
q_{t}-6rp_{x}-6pr_{x}+6qs_{x}+6sq_{x}-r_{xxx}=0 \\ 
p_{t}+6sp_{x}+6ps_{x}+6qr_{x}+6rq_{x}+s_{xxx}=0%
\end{array}%
\right. ,  \label{KdVH}
\end{equation}%
that also respects the $\mathcal{PT}_{\imath \jmath }$-symmetry. The
interesting consequence of this modification is that traveling wave
solutions $u(\xi )$ of (\ref{KdV}) depending on real combination of $x$ and $%
t$ as $\xi =x+ct\in \mathbb{R}$, with $c$ denoting the speed, become
solutions $h(\zeta )$ dependent on the hyperbolic number $\zeta =kx+ct\in 
\mathbb{D}$ instead. Interestingly a hyperbolic rotation of this number $%
\zeta $, defined as $\zeta ^{\prime }=\zeta e^{-\phi k}=kx^{\prime
}+ct^{\prime }$ with $\phi =\arctan (v/c)$, constitutes a Lorentz
transformation with $t^{\prime }=\gamma (t-v/c^{2}x)$, $x^{\prime }=\gamma
(t-vx)$ and $\gamma =1/\sqrt{1-v^{2}/c^{2}}$, see e.g. \cite%
{sobczyk,ulrych2005relativistic}.

Next we consider various solutions to these different versions of the
bicomplex KdV-equation, discuss how they may be constructed and their key
properties.

\subsubsection{One-soliton solutions with broken $\mathcal{PT}$-symmetry}

We start from the well known bright one-soliton solution of the real KdV
equation (\ref{KdV}) 
\begin{equation}
u_{\mu ,\alpha }(x,t)=\frac{\alpha ^{2}}{2}\func{sech}^{2}\left[ \frac{1}{2}%
(\alpha x-\alpha ^{3}t+\mu )\right] ,  \label{us}
\end{equation}%
when $\alpha ,\mu \mathbb{\in R}$. Since our differentials have not been
bicomplexified we may take $\mu $ to be a bicomplex number $\mu =\rho \ell
+\theta \imath +\phi \jmath +\chi k\mathbb{\in B}$ with $\rho ,\theta ,\phi
,\chi \in \mathbb{R}$, so that (\ref{us}) becomes a solution of the
bicomplex equation (\ref{KdV}). Expanding the hyperbolic function, we can
separate the bicomplex function $u_{\mu ,\alpha }(x,t)$ after some lengthy
computation into its different canonical components%
\begin{eqnarray}
u_{\rho ,\theta ,\phi ,\chi ;\alpha } &=&\frac{\ell }{2}\left[ p_{\rho +\chi
,\theta -\phi ;\alpha }+p_{\rho -\chi ,\theta +\phi ;\alpha }\right] +\frac{%
\imath }{2}\left[ q_{\rho +\chi ,\theta -\phi ;\alpha }+q_{\rho -\chi
,\theta +\phi ;\alpha }\right]   \label{unpt} \\
&&+\frac{\jmath }{2}\left[ q_{\rho -\chi ,\theta +\phi ;\alpha }-q_{\rho
+\chi ,\theta -\phi ;\alpha }\right] +\frac{k}{2}\left[ p_{\rho +\chi
,\theta -\phi ;\alpha }-p_{\rho -\chi ,\theta +\phi ;\alpha }\right] , 
\notag
\end{eqnarray}%
when using the two functions 
\begin{eqnarray}
p_{a,b;\alpha }(x,t) &=&\frac{\alpha ^{2}+\alpha ^{2}\cos b\cosh (\alpha
x-\alpha ^{3}t+a)}{\left[ \cos b+\cosh (\alpha x-\alpha ^{3}t+a)\right] ^{2}}%
,  \label{newS} \\
q_{a,b;\alpha }(x,t) &=&\frac{\alpha ^{2}\sin b\sinh (\alpha x-\alpha
^{3}t+a)}{\left[ \cos b+\cosh (\alpha x-\alpha ^{3}t+a)\right] ^{2}}.
\end{eqnarray}%
Noting that the complex solution $u_{i\theta ,\alpha }(x,t)$ studied in \cite%
{CenFring}, can be expressed as $u_{i\theta ,\alpha }(x,t)=p_{a,\theta
;\alpha }(x-a/\alpha ,t)+iq_{a,\theta ;\alpha }(x-a/\alpha ,t)$, we can also
expand the bicomplex solution (\ref{unpt}) in terms of the complex solution
as%
\begin{eqnarray}
u_{\rho ,\theta ,\phi ,\chi ;\alpha } &=&\frac{\ell }{2}\left[ u_{i(\phi
-\theta ),\alpha }\left( x+\frac{\rho +\chi }{\alpha },t\right) +u_{-i(\phi
+\theta ),\alpha }\left( x+\frac{\rho -\chi }{\alpha },t\right) \right] 
\label{uco} \\
&&+\frac{\jmath }{2}\left[ u_{-i(\phi -\theta ),\alpha }\left( x+\frac{\rho
+\chi }{\alpha },t\right) -u_{i(\phi +\theta ),\alpha }\left( x+\frac{\rho
-\chi }{\alpha },t\right) \right] .  \notag
\end{eqnarray}%
In figure \ref{Fig1} we depict the canonical components of this solution at
different times. We observe in all of them that the one-soliton solution is
split into two separate one-soliton-like components moving parallel to each
other with the same speed. The real $p$-component can be viewed as the sum
of two bright solitons and the hyperbolic $s$-component is the sum of a
bright and a dark soliton. This effect is the results of the decomposition
of each of the components into a sum of the functions $p_{a,b;\alpha }$ or $%
q_{a,b;\alpha }$, as defined in (\ref{newS}), at different values of $a$,$b$%
, but the same value of $\alpha $. Since $a$ and $b$ control the amplitude
and distance, whereas $\alpha $ regulates the speed, the constituents travel
at the same speed. We recall that this type of behaviour of degenerate
solitons can neither be created from a real nor a complex two-soliton
solution \cite{CorreaFring,CCFsineG}. So this is a novel type of phenomenon
for solitons previously not observed.

\FIGURE{ \epsfig{file=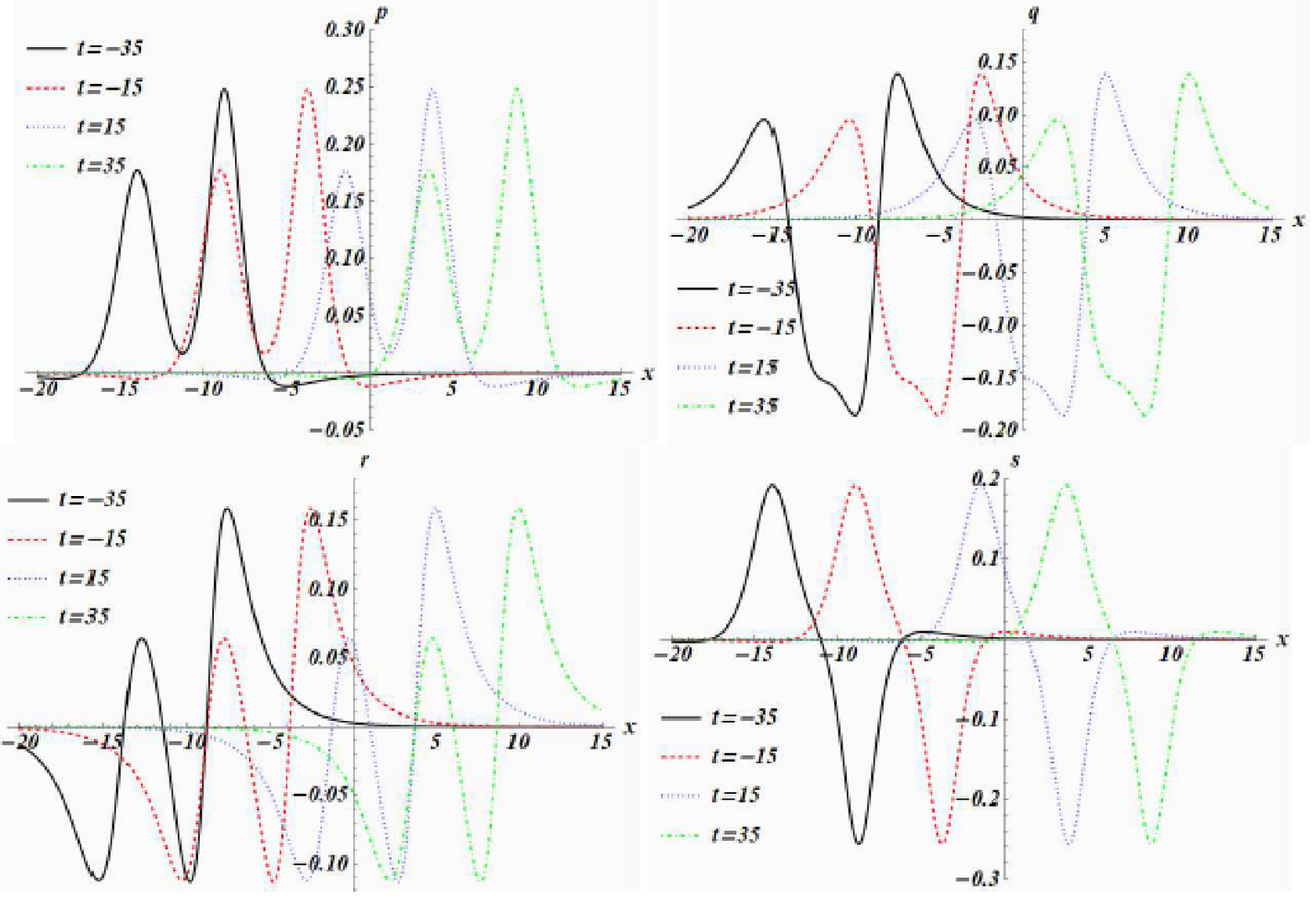,width=14.5cm}
\caption{Canonical component functions $p$, $q$, $r$ and $s$ (clockwise starting in the top left corner) of the decomposed one-soliton solution $u_{\rho ,\theta ,\phi ,\chi ;\alpha }$ 
to the bicomplex KdV equation (\ref{KdV}) with broken $\mathcal{PT}$-symmetry at different times for $\alpha =0.5
$, $\rho =1.3$, $\theta =0.4$, $\phi =2.0$ and $\chi =1.3$.}
        \label{Fig1}}

In general, the solution (\ref{us}) is not $\mathcal{PT}$-symmetric with
regard to any of the possibilities defined above. It becomes $\mathcal{PT}%
_{\imath \jmath }$-symmetric when $\rho =\chi =0$, $\mathcal{PT}_{\imath k}$%
-symmetric when $\rho =\chi =\phi =0$ and $\mathcal{PT}_{jk}$-symmetric when 
$\rho =\chi =\theta =0$.

A solution to the new KdV equation (\ref{KdVH}) is constructed as 
\begin{equation}
h_{\mu ,\alpha }(x,t)=\frac{\alpha ^{2}}{2}\func{sech}^{2}\left[ \frac{1}{2}%
(\alpha xk-\alpha ^{3}t+\mu )\right] ,
\end{equation}%
which in component form reads%
\begin{eqnarray}
h_{\rho ,\theta ,\phi ,\chi ;\alpha } &=&\frac{\ell }{2}\left[ \bar{p}_{\chi
-\rho ,\theta +\phi ;\alpha }+p_{\chi +\rho ,\theta -\phi ;\alpha }\right] +%
\frac{\imath }{2}\left[ \bar{q}_{\chi -\rho ,\theta +\phi ;\alpha }-q_{\chi
+\rho ,\theta -\phi ;\alpha }\right]  \\
&&+\frac{\jmath }{2}\left[ \bar{q}_{\chi -\rho ,\theta +\phi ;\alpha
}+q_{\chi +\rho ,\theta -\phi ;\alpha }\right] +\frac{k}{2}\left[ \bar{p}%
_{\chi +\rho ,\theta -\phi ;\alpha }-p_{\chi -\rho ,\theta +\phi ;\alpha }%
\right] ,  \notag
\end{eqnarray}%
where we introduced the notation $\bar{p}_{a,b;\alpha }(x,t)=p_{a,b;\alpha
}(x,-t)$ and $\bar{q}_{a,b;\alpha }(x,t)=q_{a,b;\alpha }(x,-t)$. 

In figure \ref{Fig2} we depict the canonical component functions of this
solution. We observe that the one-soliton solution is split into two
one-soliton-like structures that scatter head-on with each other. The real $p
$-component consists of a head-on scattering of two bright solitons and
hyperbolic the $s$-component is a head-on collision of a bright and a dark
soliton. Given that\ $u_{\rho ,\theta ,\phi ,\chi ;\alpha }$ and $h_{\rho
,\theta ,\phi ,\chi ;\alpha }(x,t)$ differ in the way that one of its
constituent functions is time-reversed this is to be expected.

\FIGURE{ \epsfig{file=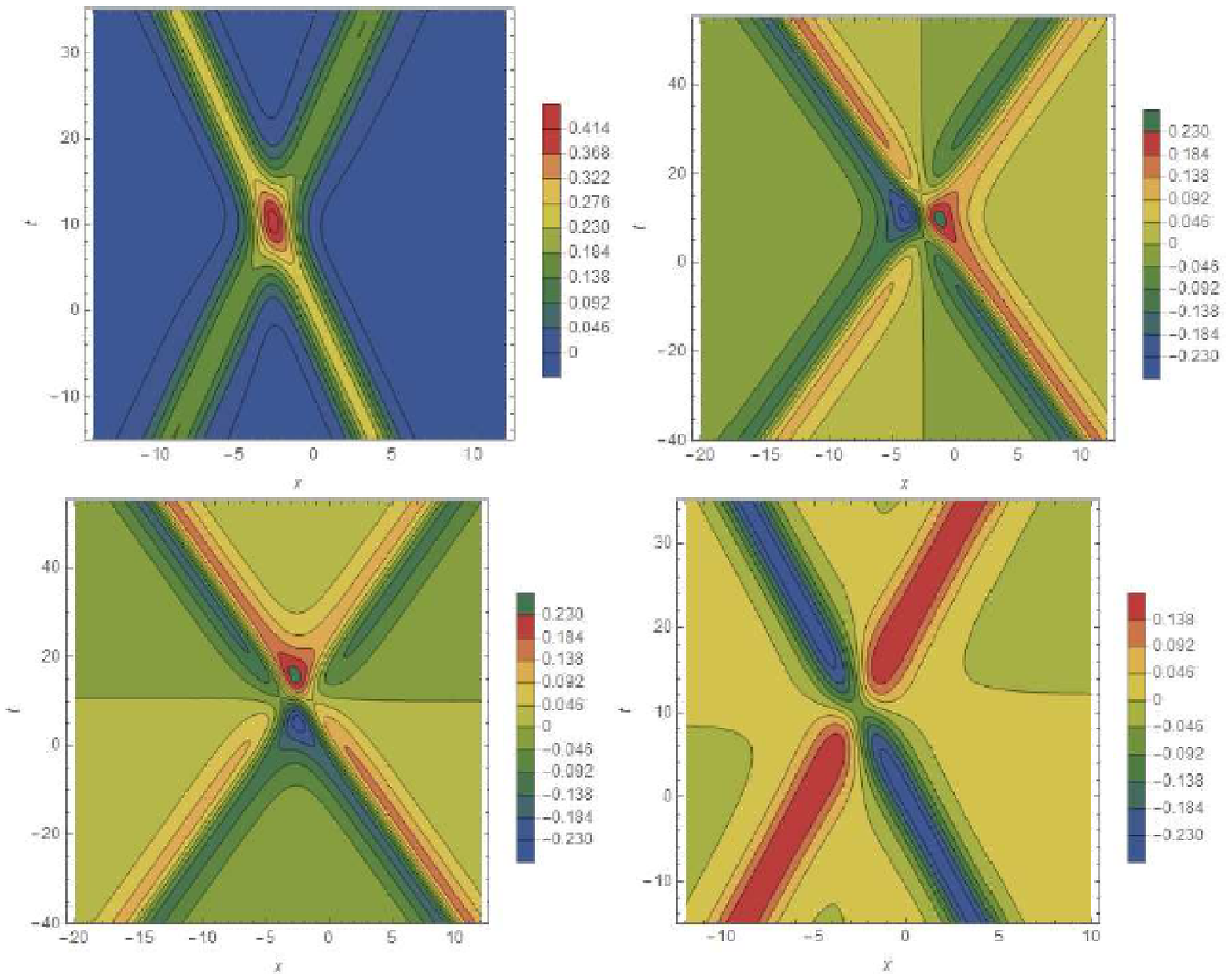,width=14.5cm}
\caption{Head-on collision of a bright soliton with a dark soliton in the canonical components $p$, $q$, $r$, $s$ (clockwise starting in the top left corner) for the one-soliton solution $h_{\rho ,\theta ,\phi ,\chi ;\alpha }$ 
to the bicomplex KdV equation (\ref{KdV}) with broken $\mathcal{PT}$-symmetry for $\alpha =0.5
$, $\rho =1.3$, $\theta =0.1$, $\phi =2.0$ and $\chi =1.3$. Time is running vertically, space horizontally and contours of the amplitudes are colour-coded indicated as in the legends.} 
        \label{Fig2}}

\subsubsection{$\mathcal{PT}_{ij}$-symmetric one-soliton solution}

\label{subptij}

An interesting solution can be constructed when we start with a complex $%
\mathcal{PT}_{\imath k}$ and a complex $\mathcal{PT}_{\jmath k}$ symmetric
solution to assemble the linear decomposition of an overall $\mathcal{PT}%
_{\imath \jmath }$-symmetric solution with different velocities. Taking in
the decomposition (\ref{SKdV}) $v(x,t)=u_{\imath \theta ,\alpha }(x,t)$ and $%
w(x,t)=u_{\jmath \phi ,\beta }(x,t)$, we can build the bicomplex
KdV-solution in the idempotent representation 
\begin{equation}
\hat{u}_{\theta ,\phi ;\alpha ,\beta }(x,t)=u_{\imath \theta ,\alpha
}(x,t)e_{1}+u_{\jmath \phi ,\beta }(x,t)e_{2}.
\end{equation}%
The expanded version in the canonical representation becomes in this case%
\begin{equation}
\hat{u}_{\theta ,\phi ;\alpha ,\beta }=\frac{\ell }{2}\left[ p_{0,\theta
;\alpha }+p_{0,\phi ;\beta }\right] +\frac{\imath }{2}\left[ q_{0,\theta
;\alpha }+q_{0,\phi ;\beta }\right] +\frac{\jmath }{2}\left[ q_{0,\phi
;\beta }-q_{0,\theta ;\alpha }\right] +\frac{k}{2}\left[ p_{0,\theta ;\alpha
}-p_{0,\phi ;\beta }\right] ,  \label{upt}
\end{equation}%
which is evidently $\mathcal{PT}_{\imath \jmath }$-symmetric. Hence this
solution contain any multicomplex shifts, but in each component two
solitonic contributions with different amplitude and speed parameter. As we
can see in figure \ref{Fig3}, in the real $p$-component a faster bright
soliton is overtaking a slower bright solitons and in hyperbolic $s$%
-component a faster bright soliton is overtaking and a slower dark soliton.
Unlike as in the real or complex case, one can carry out the limit $\beta
\rightarrow \alpha $ to the degenerate case without complication since have
the identity $\hat{u}_{\theta -\phi ,\theta +\phi ;\alpha ,\alpha
}=u_{0,\theta ,\phi ,0;\alpha }$. Similarly as in the previous section we
may also construct a further solution from a hyperbolic shift $x\rightarrow
kx$, which we do not present here.

\FIGURE{ \epsfig{file=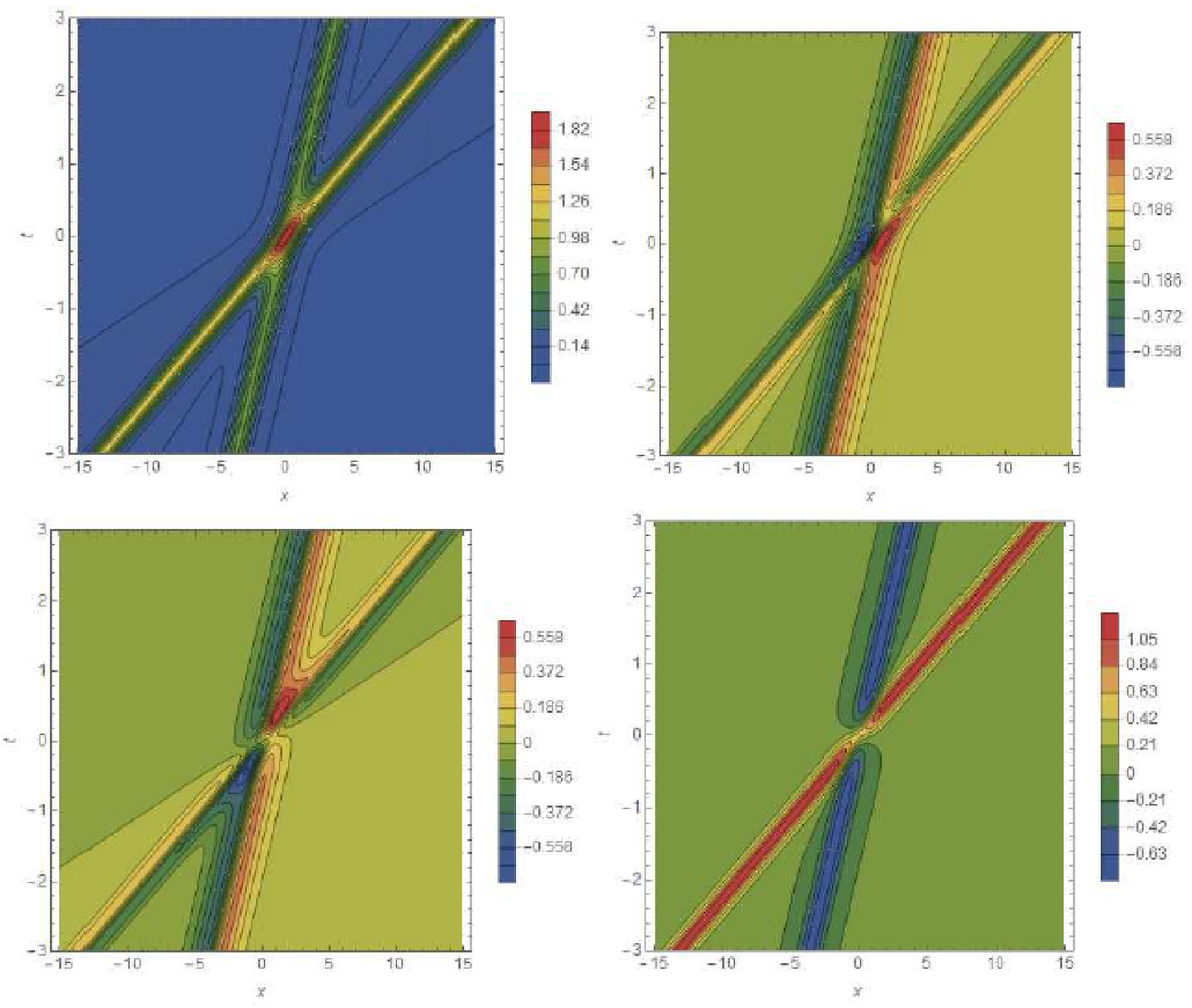,width=14.5cm} 
\caption{A fast bright soliton overtaking a slower bright soliton in the canonical component functions $p$, $q$, $r$ and $s$ (clockwise starting in the top left corner) for the one-soliton solution $\hat{u}_{\theta ,\phi ;\alpha ,\beta }$  
to the bicomplex KdV equation (\ref{KdV}) with $\mathcal{PT}_{ij}$-symmetry for $\alpha =2.1$, $\beta =1.1$, $\theta =0.6$ and $\phi =1.75$.}
        \label{Fig3}}

\subsubsection{Multi-soliton solutions}

The most compact way to express the $N$-soliton solution for the real KdV
equation in the form (\ref{KdV}) is 
\begin{equation}
u_{\mu _{1},\mu _{2},\ldots ,\mu _{n};\alpha _{1},\alpha _{2},\ldots ,\alpha
_{n}}^{(n)}(x,t)=2\left[ \ln W_{n}(\psi _{\mu _{1},\alpha _{1}},\psi _{\mu
_{2},\alpha _{2}},\ldots ,\psi _{\mu _{n},\alpha _{n}})\right] _{xx},
\label{nsol}
\end{equation}%
where $W_{n}[\psi _{1},\psi _{2},\ldots ,\psi _{n}]:=$ $\det \omega $
denotes the Wronskian with $\omega _{jk}=\partial ^{j-1}\psi _{k}/\partial
x^{j-1}$ for $j,k=1,\ldots ,n$, e.g. $W_{1}[\psi _{0}]=$ $\psi _{0}$, $%
W_{2}[\psi _{0},\psi _{1}]=$ $\psi _{0}\left( \psi _{1}\right) _{x}-\psi
_{1}\left( \psi _{0}\right) _{x}$, etc and the functions $\psi _{i}$ are
solutions to the time-independent Schr\"{o}dinger equation for the free
theory. Taking for instance $\psi _{\mu ,\alpha }(x,t)=\cosh \left[ (\alpha
x-\alpha ^{3}t+\mu )/2\right] $ for $n=1$ leads to the one-soliton solution (%
\ref{us}).

We could now take the shifts $\mu _{1},\mu _{2},\ldots ,\mu _{n}\in $ $%
\mathbb{B}$ and expand (\ref{nsol}) into its canonical components to obtain
the $N$-soliton solution for the bicomplex equation. Alternatively we may
also construct $N$-solitons in the idempotent basis in analogy to (\ref{upt}%
). We demonstrate here the latter approach for the two-soliton. From (\ref%
{nsol}) we observe that the second derivative will not alter the linear
bicomplex decomposition and it is therefore useful to introduce the quantity 
$w(x,t)$ as $u=w_{x}$. Thus a complex one-soliton solution can be obtained
from 
\begin{equation}
w_{a,b;\alpha }(x,t)=w_{a,b;\alpha }^{r}(x,t)+\imath w_{a,b;\alpha }^{i}(x,t)
\end{equation}%
with%
\begin{equation}
w_{a,b;\alpha }^{r}(x,t)=\frac{\alpha \sinh (\alpha x-\alpha ^{3}t+a)}{\cos
b+\cosh (\alpha x-\alpha ^{3}t+a)},~~~~\ w_{a,b;\alpha }^{i}(x,t)=\frac{%
\alpha \sin b}{\cos b+\cosh (\alpha x-\alpha ^{3}t+a)}.  \label{w}
\end{equation}%
Noting that $p_{a,b;\alpha }=(w_{a,b;\alpha }^{r})_{x}$, $q_{a,b;\alpha
}=(w_{a,b;\alpha }^{i})_{x}$ we obtain a complex soliton as $u_{a,b;\alpha
}=(w_{a,b;\alpha })_{x}$. Recalling now the expression%
\begin{equation}
w_{a,b,c,d;\alpha ,\beta }=\frac{\alpha ^{2}-\beta ^{2}}{w_{a,b;\alpha
}-w_{c,d;\beta }},  \label{Baeck}
\end{equation}%
from the B\"{a}cklund transformation of the complex two-soliton \cite%
{CenFring}, we can express this in terms of the functions in (\ref{w})%
\begin{equation}
w_{a,b,c,d;\alpha ,\beta }=\frac{\left( \alpha ^{2}-\beta ^{2}\right) \left[
\left( w_{a,b;\alpha }^{r}-w_{c,d;\beta }^{r}\right) -\imath \left(
w_{a,b;\alpha }^{i}-w_{c,d;\beta }^{i}\right) \right] }{\left( w_{a,b;\alpha
}^{r}-w_{c,d;\beta }^{r}\right) ^{2}+\left( w_{a,b;\alpha }^{i}-w_{c,d;\beta
}^{i}\right) ^{2}}=w_{a,b,c,d;\alpha ,\beta }^{r}+\imath w_{a,b,c,d;\alpha
,\beta }^{i}.  \label{ww}
\end{equation}%
Using (\ref{ww}) to define the two complex quantities $w_{\theta _{1},\theta
_{2},\theta _{3},\theta _{4};\alpha _{1},\alpha _{2}}=w_{2}^{r}+\imath
w_{2}^{i}~\in \mathbb{C}(\imath )$ and $w_{\phi _{1},\phi _{2},\phi
_{3},\phi _{4};\beta _{1},\beta _{2}}=\tilde{w}_{2}^{r}+\jmath ~\tilde{w}%
_{2}^{i}\in \mathbb{C}(\jmath )$ we introduce the bicomplex function 
\begin{eqnarray}
w_{2} &=&(w_{2}^{r}+\imath w_{2}^{i})e_{1}+(\tilde{w}_{2}^{r}+\jmath ~\tilde{%
w}_{2}^{i})e_{2} \\
&=&\frac{\ell }{2}\left( w_{2}^{r}+\tilde{w}_{2}^{r}\right) +\frac{\imath }{2%
}\left( w_{2}^{i}+\tilde{w}_{2}^{i}\right) +\frac{\jmath }{2}\left( \tilde{w}%
_{2}^{i}-w_{2}^{i}\right) +\frac{k}{2}\left( w_{2}^{r}-\tilde{w}%
_{2}^{r}\right) .
\end{eqnarray}%
Then by construction $u_{\theta _{1},\theta _{2},\theta _{3},\theta
_{4},\phi _{1},\phi _{2},\phi _{3},\phi _{4};\alpha _{1},\alpha _{2},\beta
_{1},\beta _{2}}=(w_{2})_{x}$ is a bicomplex two-soliton solution with four
speed parameters. In a similar fashion we can proceed to construct $N$%
-soliton for $N>2$.

\subsubsection{Real and hyperbolic conserved quantities}

Next we compute the first conserved quantities the mass $m$, the momentum $p$
and the energy $E$, see e.g. \cite{CenFring,cen2016time}%
\begin{eqnarray}
m(u) &=&\dint\nolimits_{-\infty }^{\infty }udx=m_{1}\ell +m_{2}\imath
+m_{3}\jmath +m_{4}k, \\
p(u) &=&\dint\nolimits_{-\infty }^{\infty }u^{2}dx=p_{1}\ell +p_{2}\imath
+p_{3}\jmath +p_{4}k, \\
E(u) &=&\dint\nolimits_{-\infty }^{\infty }\left( \frac{1}{2}%
u_{x}^{2}-u^{3}\right) dx=E_{1}\ell +E_{2}\imath +E_{3}\jmath +E_{4}k
\end{eqnarray}%
Decomposing the relevant densities into the canonical basis, $u$ as in (\ref%
{ub}), $u^{2}$ as 
\begin{equation}
u^{2}=\left( p^{2}-q^{2}-r^{2}+s^{2}\right) \ell +2(pq-rs)\imath
+2(pr-qs)\jmath +2(qr+ps)k
\end{equation}%
and the Hamiltonian density $\mathcal{H}(u,u_{x})=u_{x}^{2}/2-u^{3}$ as 
\begin{eqnarray}
\mathcal{H} &=&\left[ 3p\left( q^{2}+r^{2}-s^{2}\right) +\frac{%
p_{x}^{2}-q_{x}^{2}-r_{x}^{2}+s_{x}^{2}}{2}-6qrs-p^{3}\right] \ell  \\
&&+\left[ q^{3}-3p^{2}q+p_{x}q_{x}+6prs+3q\left( r^{2}-s^{2}\right)
-r_{x}s_{x}\right] \imath   \notag \\
&&+\left[ r^{3}+6pqs+3r\left( q^{2}-s^{2}-p^{2}\right) +p_{x}r_{x}-q_{x}s_{x}%
\right] \jmath   \notag \\
&&+\left[ 3s\left( r^{2}-p^{2}+q^{2}\right) -6pqr+p_{x}s_{x}+q_{x}r_{x}-s^{3}%
\right] k,  \notag
\end{eqnarray}%
we integrate componentwise. For the solutions $u_{\rho ,\theta ,\phi ,\chi
;\alpha }$ and $h_{\rho ,\theta ,\phi ,\chi ;\alpha }$ with broken $\mathcal{%
PT}$-symmetry we obtain the real conserved quantities 
\begin{eqnarray}
m(u_{\rho ,\theta ,\phi ,\chi ;\alpha }) &=&m(h_{\rho ,\theta ,\phi ,\chi
;\alpha })=2\alpha \ell ,  \label{mr1} \\
p(u_{\rho ,\theta ,\phi ,\chi ;\alpha }) &=&p(h_{\rho ,\theta ,\phi ,\chi
;\alpha })=\frac{2}{3}\alpha ^{3}\ell ,  \label{mr2} \\
E(u_{\rho ,\theta ,\phi ,\chi ;\alpha }) &=&E(h_{\rho ,\theta ,\phi ,\chi
;\alpha })=-\frac{1}{5}\alpha ^{5}\ell .  \label{mr3}
\end{eqnarray}%
These values are the same as those found in \cite{CenFring} for the complex
solitons. Given that the $\mathcal{PT}$-symmetries are all broken this is
surprising at first sight. However, considering the representation (\ref{uco}%
) this is easily understood when using the result of \cite{CenFring}. Then $%
m(u_{\rho ,\theta ,\phi ,\chi ;\alpha })$ is simply $\ell /2(2\alpha
+2\alpha )+\jmath /2(2\alpha -2\alpha )=2\alpha \ell $. We can argue
similarly for the other conserved quantities.

For the $\mathcal{PT}_{ij}$-symmetric solution $\hat{u}_{\theta ,\phi
;\alpha ,\beta }$ we obtain the following hyperbolic values for the
conserved quantities 
\begin{eqnarray}
m(\hat{u}_{\theta ,\phi ;\alpha ,\beta }) &=&(\alpha +\beta )\ell +(\alpha
-\beta )k, \\
p(\hat{u}_{\theta ,\phi ;\alpha ,\beta }) &=&\frac{1}{3}\left( \alpha
^{3}+\beta ^{3}\right) \ell +\frac{1}{3}\left( \alpha ^{3}-\beta ^{3}\right)
k \\
E(\hat{u}_{\theta ,\phi ;\alpha ,\beta }) &=&-\left( \frac{\alpha ^{5}}{10}+%
\frac{\beta ^{5}}{10}\right) \ell +\left( \frac{\beta ^{5}}{10}-\frac{\alpha
^{5}}{10}\right) k.
\end{eqnarray}%
The values become real and coincide with the expressions (\ref{mr1})-(\ref%
{mr3}) when we sum up the contributions from the real and hyperbolic
component or in the degenerate case when we take the limit $\beta
\rightarrow \alpha $.

\subsection{The bicomplex Alice and Bob KdV equation}

Various nonlocal versions of nonlinear wave equations that have been
overlooked previously have attracted considerable attention recently. In
reference to standard scenarios in quantum cryptography some of them are
also often referred to as Alice and Bob systems. These variants of the
nonlinear Schr\"{o}dinger or Hirota equation \cite%
{ablowitz2013,CenFringHir,stalin2017,manikandan2018dynamical} arise from an
alternative choice in the compatibility condition of the two AKNS-equations.
For the KdV equation (\ref{KdV}) they can be constructed \cite%
{AliceB1,AliceB2,AliceB3} by choosing $u(x,t)=1/2\left[ a(x,t)+b(x,t)\right] 
$, with the constraint $\mathcal{PT}a(x,t)=a(-x,-t)=b(x,t)$, thus converting
it into an equation that can be decomposed into two equations, the Alice and
Bob KdV (ABKdV) equation%
\begin{eqnarray}
a_{t}+3/4(a+b)(3a_{x}+b_{x})+a_{xxx} &=&0,  \label{AB1} \\
b_{t}+3/4(a+b)(a_{x}+3b_{x})+b_{xxx} &=&0.  \label{AB2}
\end{eqnarray}%
In a similar way as the two AKNS-equations can be made compatible by a
suitable transformation map, these two equations are converted into each
other by a $\mathcal{PT}$-transformation, i.e. $\mathcal{PT}$(\ref{AB1})$%
\equiv $(\ref{AB2}). Evidently the decomposition is not unique and one may
also add and subtract a constrained function of $a$ and $b$ or consider
different types of maps to relate the equation.

The bicomplex version of the Alice and Bob system (\ref{AB1}), (\ref{AB2})
is obtained by taking $a,b\in \mathbb{B}$. In the canonical basis we use the
conventions $u(x,t)=\ell p(x,t)+\imath q(x,t)+\jmath r(x,t)+ks(x,t)$, $%
a(x,t)=\ell \hat{p}(x,t)+\imath \hat{q}(x,t)+\jmath \hat{r}(x,t)+k\hat{s}%
(x,t)$, $b(x,t)=\ell \check{p}(x,t)+\imath \check{q}(x,t)+\jmath \check{r}%
(x,t)+k\check{s}(x,t)$, so that the ABKdV equations (\ref{AB1}) and (\ref%
{AB2}) decompose into eight coupled equations%
\begin{eqnarray}
\hat{p}_{t} &=&-\hat{p}_{xxx}-\frac{3}{2}\left[ p\left( \check{p}_{x}+3\hat{p%
}_{x}\right) -q\left( \check{q}_{x}+3\hat{q}_{x}\right) -r\left( \check{r}%
_{x}+3\hat{r}_{x}\right) +s\left( \check{s}_{x}+3\hat{s}_{x}\right) \right] ,
\label{AB81} \\
\hat{q}_{t} &=&-\hat{q}_{xxx}+\frac{3}{2}\left[ p\left( \check{q}_{x}+3\hat{q%
}_{x}\right) +q\left( \check{p}_{x}+3\hat{p}_{x}\right) -r\left( \check{s}%
_{x}+3\hat{s}_{x}\right) -s\left( \check{r}_{x}+3\hat{r}_{x}\right) \right] ,
\\
\hat{r}_{t} &=&-\hat{r}_{xxx}+\frac{3}{2}\left[ p\left( \check{r}_{x}+3\hat{r%
}_{x}\right) -q\left( \check{s}_{x}+3\hat{s}_{x}\right) +r\left( \check{p}%
_{x}+3\hat{p}_{x}\right) -s\left( \check{q}_{x}+3\hat{q}_{x}\right) \right] ,
\\
\hat{s}_{t} &=&-\hat{s}_{xxx}+\frac{3}{2}\left[ p\left( \check{s}_{x}+3\hat{s%
}_{x}\right) +q\left( \check{r}_{x}+3\hat{r}_{x}\right) +r\left( \check{q}%
_{x}+3\hat{q}_{x}\right) +s\left( \check{p}_{x}+3\hat{p}_{x}\right) \right] ,
\end{eqnarray}%
and%
\begin{eqnarray}
\check{p}_{t} &=&-\check{p}_{xxx}+\frac{3}{2}\left[ p\left( 3\check{p}_{x}+%
\hat{p}_{x}\right) -q\left( 3\check{q}_{x}+\hat{q}_{x}\right) -r\left( 3%
\check{r}_{x}+\hat{r}_{x}\right) +s\left( 3\check{s}_{x}+\hat{s}_{x}\right) %
\right] ,  \label{AB85} \\
\check{q}_{t} &=&-\check{q}_{xxx}+\frac{3}{2}\left[ p\left( 3\check{q}_{x}+%
\hat{q}_{x}\right) +q\left( 3\check{p}_{x}+\hat{p}_{x}\right) -r\left( 3%
\check{s}_{x}+\hat{s}_{x}\right) -s\left( 3\check{r}_{x}+\hat{r}_{x}\right) %
\right] , \\
\check{r}_{t} &=&-\check{r}_{xxx}+\frac{3}{2}\left[ p\left( 3\check{r}_{x}+%
\hat{r}_{x}\right) -q\left( 3\check{s}_{x}+\hat{s}_{x}\right) +r\left( 3%
\check{p}_{x}+\hat{p}_{x}\right) -s\left( 3\check{q}_{x}+\hat{q}_{x}\right) %
\right] , \\
\check{s}_{t} &=&-\check{s}_{xxx}+\frac{3}{2}\left[ p\left( 3\check{s}_{x}+%
\hat{s}_{x}\right) +q\left( 3\check{r}_{x}+\hat{r}_{x}\right) +r\left( 3%
\check{q}_{x}+\hat{q}_{x}\right) +s\left( 3\check{p}_{x}+\hat{p}_{x}\right) %
\right] .  \label{AB88}
\end{eqnarray}

A real solution to the ABKdV equations (\ref{AB1}) and (\ref{AB2}) that sums
up to the standard one-soliton solution (\ref{us}) is found as%
\begin{eqnarray}
a_{\mu ,\nu ;\alpha }(x,t) &=&u_{\mu ,\alpha }(x,t)+\nu \tanh \left[ \frac{1%
}{2}(\alpha x-\alpha ^{3}t+\mu )\right] ,\qquad \\
b_{\mu ,\nu ;\alpha }(x,t) &=&u_{\mu ,\alpha }(x,t)-\nu \tanh \left[ \frac{1%
}{2}(\alpha x-\alpha ^{3}t+\mu )\right] ,
\end{eqnarray}%
with arbitrary constants $\nu ,\mu \in \mathbb{R}$. Proceeding now as for
the local variant by taking $\mu =\rho \ell +\theta \imath +\phi \jmath
+\chi k\mathbb{\in B}$, we decompose $a_{\mu ,\nu ;\alpha }$ and $b_{\mu
,\nu ;\alpha }$ into their canonical components and obtain after some
lengthy computation the corresponding solution to the bicomplex version of
the ABKdV equations (\ref{AB81})-(\ref{AB88}) as%
\begin{equation}
a_{\rho ,\theta ,\phi ,\chi ;\alpha }=u_{\rho ,\theta ,\phi ,\chi ;\alpha }+%
\frac{\ell }{2}\nu F_{w+\rho ,\theta ,\phi ,\chi }+\frac{\imath }{2}\nu
G_{w+\rho ,\theta ,\phi ,\chi }+\frac{\jmath }{2}\nu G_{w+\rho ,\phi ,\theta
,\chi }+\frac{k}{2}\nu F_{\chi ,\theta ,\phi ,w+\rho }  \label{a1}
\end{equation}%
with $w_{\alpha }=\alpha x-\alpha ^{3}t$ and the newly defined functions%
\begin{eqnarray}
F_{x_{1},x_{2},x_{3},x_{4}}\! &=&\!\frac{\sinh \text{$x_{1}$}\sec \text{$%
x_{2}$}\sec \text{$x_{3}$}\func{sech}\text{$x_{4}$}+\tanh \text{$x_{1}$}%
-\tan \text{$x_{2}$}\tan \text{$x_{3}$}\tanh \text{$x_{4}$}}{1-\tanh \text{$%
x_{1}$}\tan \text{$x_{2}$}\tan \text{$x_{3}$}\tanh \text{$x_{4}$}+\frac{%
\cosh (2\text{$x_{1}$})+\cos (2\text{$x_{2}$})+\cos (2\text{$x_{3}$})+\cosh
(2\text{$x_{4}$})}{4\cosh \text{$x_{1}$}\cos \text{$x_{2}$}\cos \text{$x_{3}$%
}\cosh \text{$x_{4}$}}},~~~~~~~~ \\
G_{x_{1},x_{2},x_{3},x_{4}}\! &=&\!\frac{\func{sech}\text{$x_{1}$}\sec \text{%
$x_{2}$}\sin \text{$x_{3}$}\func{sech}\text{$x_{4}$}+\tan \text{$x_{3}$}%
+\tanh \text{$x_{1}$}\tan \text{$x_{2}$}\tanh \text{$x_{4}$}}{1-\tanh \text{$%
x_{1}$}\tan \text{$x_{2}$}\tan \text{$x_{3}$}\tanh \text{$x_{4}$}+\frac{%
\cosh (2\text{$x_{1}$})+\cos (2\text{$x_{2}$})+\cos (2\text{$x_{3}$})+\cosh
(2\text{$x_{4}$})}{4\cosh \text{$x_{1}$}\cos \text{$x_{2}$}\cos \text{$x_{3}$%
}\cosh \text{$x_{4}$}}}.
\end{eqnarray}%
The functions $b_{\rho ,\theta ,\phi ,\chi ;\alpha }$, or equivalently the
individual components $\check{p},\check{q},\check{r},\check{s}$, are
obtained by a $\mathcal{PT}$-transformation.

We may also proceed as in subsection \ref{subptij} and construct a solution
in the idempotent representation. Keeping the parameter $\nu $ real, a
solution based on the idempotent decomposition is%
\begin{eqnarray}
a_{\theta ,\phi ,\nu ;\alpha ,\beta } &=&a_{i\theta ,\nu ;\alpha
}e_{1}+a_{i\phi ,\nu ;\beta }e_{2}  \label{twos} \\
&=&\hat{u}_{\theta ,\phi ;\alpha ,\beta }+\frac{\ell }{2}\nu (F_{w_{\alpha
},\theta ,0,0}+F_{w_{\beta },\phi ,0,0})+\frac{\imath }{2}\nu (G_{w_{\alpha
},0,\theta ,0}+G_{w_{\beta },0,\phi ,0})  \notag \\
&&+\frac{\jmath }{2}\nu \left( G_{w_{\beta },0,\phi ,0}+G_{w_{\alpha
},0,\theta ,0}\right) +\frac{k}{2}\nu \left( F_{w_{\alpha },\theta
,0,0}-F_{w_{\beta },\phi ,0,0}\right) .  \notag
\end{eqnarray}%
Once more, the functions $b_{\rho ,\theta ,\phi ,\chi ;\alpha }$ or $\check{p%
},\check{q},\check{r},\check{s}$ are obtained by a $\mathcal{PT}$%
-transformation. Comparing (\ref{twos}) with $a_{\rho ,\theta ,\phi ,\chi
;\alpha }$ in (\ref{a1}) we have now two speed parameters at our disposal,
similarly as in the local case.

\section{Quaternionic solitons}

\subsection{Quaternionic numbers and functions}

The quaternions in the canonical basis are defined as the set of elements%
\begin{equation}
\mathbb{H}=\left\{ a_{1}\ell +a_{2}\imath +a_{3}\jmath
+a_{4}k~|~a_{1},a_{2},a_{3},a_{4}\in \mathbb{R}\right\} .
\end{equation}%
The multiplication of the basis $\{\ell ,\imath ,\jmath ,k\}$ is
noncommutative with $\ell $ denoting the real unit element, $\ell ^{2}=1$
and $\imath ,\jmath ,k$ its three imaginary units with $\imath ^{2}=\jmath
^{2}=k^{2}=-1$. The remaining multiplication rules are $\imath \jmath
=-\jmath \imath =k$, $\jmath k=-k\jmath =\imath $ and $k\imath =-\imath
k=\jmath $. The multiplication table remains invariant under the symmetries $%
\mathcal{PT}_{\imath \jmath }$, $\mathcal{PT}_{\imath k}$ and $\mathcal{PT}%
_{\jmath k}$. Using these rules for the basis, two quaternions in the
canonical basis $n_{a}=a_{1}\ell +a_{2}\imath +a_{3}\jmath +a_{4}k\in 
\mathbb{H}$ and $n_{b}=b_{1}\ell +b_{2}\imath +b_{3}\jmath +b_{4}k\in 
\mathbb{H}$ are multiplied as%
\begin{eqnarray}
n_{a}n_{b} &=&\left( a_{1}b_{1}-a_{2}b_{2}-a_{3}b_{3}-a_{4}b_{4}\right) \ell
+\left( a_{1}b_{2}+a_{2}b_{1}+a_{3}b_{4}-a_{4}b_{3}\right) \imath 
\label{m2} \\
&&+\left( a_{1}b_{3}-a_{2}b_{4}+a_{3}b_{1}+a_{4}b_{2}\right) \jmath +k\left(
a_{1}b_{4}+a_{2}b_{3}-a_{3}b_{2}+a_{4}b_{1}\right) k.  \notag
\end{eqnarray}%
There are various representations for quaternions, see e.g. \cite%
{sangwine2011fundamental}, of which the complex form will be especially
useful for what follows. With the help of (\ref{m2}) one easily verifies that%
\begin{equation}
\xi :=\frac{1}{\mathcal{N}}\left( a_{2}\imath +a_{3}\jmath +a_{4}k\right)
\quad \text{with }\mathcal{N=}\sqrt{a_{2}^{2}+a_{3}^{2}+a_{4}^{2}}
\label{ima}
\end{equation}%
constitutes a new imaginary unit with\ $\xi ^{2}=-1$. This means that in
this representation we can formally view a quaternion, $n_{a}\in \mathbb{H}$%
, as an element in the complex numbers%
\begin{equation}
n_{a}=a_{1}\ell +\xi \mathcal{N\in \mathbb{C}(\xi ),}  \label{cf}
\end{equation}%
with real part $a_{1}$ and imaginary part $\mathcal{N}$. Notice that a $%
\mathcal{PT}_{\xi }$-symmetry can only be achieved with a $\mathcal{PT}%
_{\imath \jmath k}$-symmetry acting on the unit vectors in the canonical
representation. Unlike the bicomplex numbers or the coquaternions, see
below, the quaternionic algebra does not contain any idempotents.

\subsection{The quaternionic Korteweg-de Vries equation}

Applying now the multiplication law (\ref{m2}) to quaternionic functions,
the KdV equation for a quaternionic field of the form $u(x,t)=\ell
p(x,t)+\imath q(x,t)+\jmath r(x,t)+ks(x,t)\in \mathbb{H}$ can also be viewed
as a set of coupled equations for the four real fields $p(x,t)$, $q(x,t)$, $%
r(x,t)$, $s(x,t)\in \mathbb{R}$ 
\begin{equation}
u_{t}+6uu_{x}+u_{xxx}=0\quad \Leftrightarrow \quad \left\{ 
\begin{array}{r}
p_{t}+6pp_{x}-6qq_{x}-6rr_{x}-6ss_{x}+p_{xxx}=0 \\ 
q_{t}+6qp_{x}+6pq_{x}-6sr_{x}+6rs_{x}+q_{xxx}=0 \\ 
r_{t}+6rp_{x}+6pr_{x}-6qs_{x}+6sq_{x}+r_{xxx}=0 \\ 
s_{t}+6sp_{x}+6ps_{x}+6qr_{x}-6rq_{x}+s_{xxx}=0%
\end{array}%
\right. .  \label{qKdV}
\end{equation}%
Notice that when comparing the bicomplex KdV equation (\ref{KdV}) and the
quaternionic\ KdV equation (\ref{qKdV}) only the signs of the penultimate
terms in all four equations have changed. This means that also (\ref{qKdV})
is invariant under the $\mathcal{PT}_{\imath \jmath }$-symmetry.
Alternatively, we may consider here the aforementioned symmetry 
\begin{equation}
\mathcal{PT}_{\imath \jmath k}:x\rightarrow -x,\ t\rightarrow -t,\imath
\rightarrow -\imath ,\jmath \rightarrow -\jmath ,k\rightarrow
-k,p\rightarrow p,q\rightarrow -q,r\rightarrow -r,s\rightarrow
-s,u\rightarrow u,
\end{equation}%
which violates all the noncommutative multiplication rules $\imath \jmath
=-\jmath \imath =k$, $\jmath k=-k\jmath =\imath $ and $k\imath =-\imath
k=\jmath $. Thus in order to implement the symmetry $\mathcal{PT}_{\imath
\jmath k}$ we must set all terms resulting from these multiplications to
zero, so that we obtain the additional constraints%
\begin{equation}
sr_{x}=rs_{x},\qquad qs_{x}=sq_{x},\qquad \text{and\qquad }qr_{x}=rq_{x}.
\label{conq}
\end{equation}%
When eliminating these terms from (\ref{qKdV}) the remaining set of
equations is $\mathcal{PT}_{\imath \jmath k}$-symmetric, which appears to be
a rather strong imposition. However, the equations without these terms
emerge quite naturally when keeping in mind that the product of functions in
(\ref{qKdV}) is noncommutative so that one should symmetrize products and
replace $6uu_{x}\rightarrow 3uu_{x}+3u_{x}u$. This process corresponds
precisely to imposing the constraints (\ref{conq}).

\subsection{$\mathcal{PT}_{\imath \jmath k}$-symmetric N-soliton solutions 
\label{comsol}}

Due to the noncommutative nature of the quaternions it appears difficult at
first sight to find solutions to the quaternionic KdV equation. However,
using the complex representation (\ref{cf}), and imposing the $\mathcal{PT}%
_{\imath \jmath k}$-symmetric, we may resort to our previous analysis on
complex solitons. Following \cite{CenFring} and considering the shifted
solution (\ref{us}) in the complex space $\mathbb{C}(\xi )$ yields the
solution%
\begin{eqnarray}
u_{a_{1}\ell +\xi \mathcal{N},\alpha }(x,t) &=&p_{a_{1},\mathcal{N};\alpha
}(x,t)-\xi q_{a_{1},\mathcal{N};\alpha }(x,t)  \label{sq1} \\
&=&p_{a_{1},\mathcal{N};\alpha }(x,t)\ell -\frac{1}{\mathcal{N}}q_{a_{1},%
\mathcal{N};\alpha }(x,t)\left( a_{2}\imath +a_{3}\jmath +a_{4}k\right) .
\label{sq2}
\end{eqnarray}%
This solution becomes $\mathcal{PT}_{\imath \jmath k}$-symmetric when we
carry out a shift in $x$ or $t$ to eliminate the real part of the shift.
Reading off the functions $p(x,t)$, $q(x,t)$, $r(x,t)$, $s(x,t)$ from (\ref%
{sq2}), it is also obvious that the constraints (\ref{conq}) are indeed
satisfied. Thus the real $\ell $-component is a one-solitonic structure
similar to the real part of a complex soliton and the remaining component
consists of the imaginary parts of a complex soliton with overall different
amplitudes. It is clear that the conserved quantities constructed from this
solution must be real, which follows by using the same argument as for the
imaginary part in the complex case \cite{CenFring} separately for each of
the $\imath $,$\jmath $,$k$-components. By considering all functions to be
in $\mathbb{C}(\xi )$, it is also clear that multi-soliton solutions can be
constructed in analogy to the complex case $\mathbb{C}(\imath )$ treated in 
\cite{CenFring} with a subsequent expansion into canonical components.

Since the quaternionic algebra does not contain any idempotents, a
construction similar to the one carried out in subsection \ref{subptij} does
not seem to be possible for quaternions. However, we can use (\ref{Baeck})
for two complex solutions $w_{a,b;\alpha }(x,t)=w_{a,b;\alpha }^{r}(x,t)+\xi
_{\alpha }w_{a,b;\alpha }^{i}(x,t)$, $w_{c,d;\beta }(x,t)=w_{c,d;\beta
}^{r}(x,t)+\xi _{b}w_{c,d;\beta }^{i}(x,t)$, where the imaginary units are
defined as in (\ref{ima}) with $\xi _{a}(a_{2},a_{3},a_{4})$ and $\xi
_{b}(b_{2},b_{3},b_{4})$. Expanding that expression in the canonical basis
we obtain 
\begin{equation}
w_{2}=\frac{\alpha ^{2}-\beta ^{2}}{\omega _{1}^{2}+\omega _{2}^{2}+\omega
_{3}^{2}+\omega _{4}^{2}}\left( \ell \omega _{1}-\imath \omega _{2}-\jmath
\omega _{3}-k\omega _{4}\right)  \label{wco}
\end{equation}%
with%
\begin{equation}
\omega _{1}=w_{a,b;\alpha }^{r}-w_{c,d;\beta }^{r},\quad \omega _{n}=\frac{%
a_{n}w_{a,b;\alpha }^{i}}{\mathcal{N}_{a}}-\frac{b_{n}w_{c,d;\beta }^{i}}{%
\mathcal{N}_{b}},~~~~n=2,3,4.
\end{equation}%
A coquaternionic two-soliton solution to (\ref{qKdV}) is then obtained from (%
\ref{wco}) as $u^{(2)}=(w_{2})_{x}$.

\section{Coquaternionic solitons}

\subsection{Coquaternionic numbers and functions}

The coquaternions or often also referred to as split-quaternions in the
canonical basis are defined as the set of elements%
\begin{equation}
\mathbb{P}=\left\{ a_{1}\ell +a_{2}\imath +a_{3}\jmath
+a_{4}k~|~a_{1},a_{2},a_{3},a_{4}\in \mathbb{R}\right\} .  \label{co1}
\end{equation}%
The multiplication of the basis $\{\ell ,\imath ,\jmath ,k\}$ is
noncommutative with a real unit element $\ell $, $\ell ^{2}=1$, two
hyperbolic unit elements $\jmath ,k$, $\jmath ^{2}=k^{2}=1$, and one
imaginary unit $\imath ^{2}=-1$. The remaining multiplication rules are $%
\imath \jmath =-\jmath \imath =k$, $\jmath k=-k\jmath =-\imath $ and $%
k\imath =-\imath k=\jmath $. The multiplication table remains invariant
under the symmetries $\mathcal{PT}_{\imath \jmath }$, $\mathcal{PT}_{\imath
k}$ and $\mathcal{PT}_{\jmath k}$. Using these rules for the basis, two
coquaternions in the canonical basis $n_{a}=a_{1}\ell +a_{2}\imath
+a_{3}\jmath +a_{4}k\in \mathbb{P}$ and $n_{b}=b_{1}\ell +b_{2}\imath
+b_{3}\jmath +b_{4}k\in \mathbb{P}$ are multiplied as%
\begin{eqnarray}
n_{a}n_{b} &=&\left( a_{1}b_{1}-a_{2}b_{2}+a_{3}b_{3}+a_{4}b_{4}\right) \ell
+\left( a_{1}b_{2}+a_{2}b_{1}-a_{3}b_{4}+a_{4}b_{3}\right) \imath  \label{m3}
\\
&&+\left( a_{1}b_{3}-a_{2}b_{4}+a_{3}b_{1}+a_{4}b_{2}\right) \jmath +k\left(
a_{1}b_{4}+a_{2}b_{3}-a_{3}b_{2}+a_{4}b_{1}\right) k.  \notag
\end{eqnarray}%
There are various coquaternionic representations for numbers and functions.
Similar as a quaternion one can formally view a coquaternion, $n_{1}\in 
\mathbb{P}$, as an element in the complex numbers%
\begin{equation}
n_{a}=a_{1}\ell +\zeta \mathcal{M\in \mathbb{C}(\zeta )}  \label{cocompl}
\end{equation}%
with real part $a_{1}$ and imaginary part $\mathcal{M}$. The new imaginary
unit, $\zeta ^{2}=-1$, 
\begin{equation}
\zeta :=\frac{1}{\mathcal{M}}\left( a_{2}\imath +a_{3}\jmath +a_{4}k\right)
\quad \text{with }\mathcal{M=}\sqrt{a_{2}^{2}-a_{3}^{2}-a_{4}^{2}}
\end{equation}%
is, however, only defined for $a_{2}^{2}\neq a_{3}^{2}+a_{4}^{2}$. For
definiteness we assume here $\left\vert a_{2}\right\vert >\sqrt{%
a_{3}^{2}+a_{4}^{2}}$. Similarly as the $\mathcal{PT}_{\xi }$-symmerty also
the $\mathcal{PT}_{\zeta }$-symmerty requires a $\mathcal{PT}_{\imath \jmath
k}$-symmetry. Unlike the quaternions, the coquaternions possess a number
idempotents $e_{1}=(1+k)/2$, $e_{2}=(1-k)/2$ with $e_{1}^{2}=e_{1}$, $%
e_{2}^{2}=e_{2}$, $e_{1}e_{2}=0$ or $e_{3}=(1+\jmath )/2$, $e_{4}=(1-\jmath
)/2$ with $e_{3}^{2}=e_{3}$, $e_{4}^{2}=e_{4}$, $e_{3}e_{4}=0$. So for
instance, $n_{a}$ is an element in 
\begin{equation}
\mathbb{P}=\left\{ e_{1}v_{1}+e_{2}v_{2}~|~v_{1}\in \mathbb{D}(\jmath
),v_{2}\in \mathbb{D}(\jmath )\right\} ,  \label{idco}
\end{equation}%
where the hyperbolic numbers in (\ref{idco}) are related to the coefficient
in the canonical basis as $v_{1}=(a_{1}+a_{4})\ell +(a_{2}+a_{3})\jmath $
and $v_{2}=(a_{1}-a_{4})\ell +(a_{3}-a_{2})\jmath $.

\subsection{The coquaternionic Korteweg-de Vries equation}

Applying now the multiplication law (\ref{m3}) to coquaternionic functions,
the KdV equation for a quaternionic field of the form $u(x,t)=\ell
p(x,t)+\imath q(x,t)+\jmath r(x,t)+ks(x,t)\in \mathbb{P}$ can also be viewed
as a set of coupled equations for the four real fields $p(x,t)$, $q(x,t)$, $%
r(x,t)$, $s(x,t)\in \mathbb{R}$. The symmetric coquaternionic KdV equation
then becomes 
\begin{equation}
u_{t}+3(uu_{x}+u_{x}u)+u_{xxx}=0\quad \Leftrightarrow \quad \left\{ 
\begin{array}{r}
p_{t}+6pp_{x}-6qq_{x}+6ss_{x}+6rr_{x}+p_{xxx}=0 \\ 
q_{t}+6qp_{x}+6pq_{x}+q_{xxx}=0 \\ 
r_{t}+6rp_{x}+6pr_{x}+r_{xxx}=0 \\ 
s_{t}+6sp_{x}+6ps_{x}+s_{xxx}=0%
\end{array}%
\right. .  \label{KdVco}
\end{equation}%
Notice that the last three equations of the coupled equation in (\ref{KdVco}%
) are identical to the symmetric quaternionic KdV equation (\ref{qKdV}) with
constraints (\ref{conq}).

\subsection{$\mathcal{PT}_{\imath \jmath k}$-symmetric N-soliton solutions}

Using the representation (\ref{cocompl}) we proceed as in subsection \ref%
{comsol} and consider the shifted solution (\ref{us}) in the complex space $%
\mathbb{C}(\zeta )$ 
\begin{eqnarray}
u_{a_{1}\ell +\zeta \mathcal{M},\alpha }(x,t) &=&p_{a_{1},\mathcal{M};\alpha
}(x,t)-\zeta q_{a_{1},\mathcal{M};\alpha }(x,t)  \label{soco2} \\
&=&p_{a_{1},\mathcal{M};\alpha }(x,t)\ell -\frac{1}{\mathcal{M}}q_{a_{1},%
\mathcal{M};\alpha }(x,t)\left( a_{2}\imath +a_{3}\jmath +a_{4}k\right)
\end{eqnarray}%
that solves the\ coquaternionic KdV equation (\ref{KdVco}). The solution in (%
\ref{soco2}) is $\mathcal{PT}_{\imath \jmath k}$-symmetric. Multi-soliton
solutions can be constructed in analogy to the complex case $\mathbb{C}%
(\imath )$ treated in \cite{CenFring} by treating all functions in $\mathbb{C%
}(\zeta )$ as explained in more detail at the end of section 4.

\section{Octonionic solitons}

We finish our discussion with a comment on the construction of octonionic
solitons. Octonions or Cayley numbers are extensions of the quaternions with
a doubling of the dimensions. In the canonical basis they can be represented
as%
\begin{equation}
\mathbb{O}=\left\{
a_{0}e_{0}+a_{1}e_{1}+a_{2}e_{2}+a_{3}e_{3}+a_{4}e_{4}+a_{5}e_{5}+a_{6}e_{6}+a_{7}e_{7}|~a_{i}\in 
\mathbb{R}\right\} .
\end{equation}%
The multiplication of the units is defined by noting that each of the seven
quadruplets $(e_{0},e_{1},e_{2},e_{3})$, $(e_{0},e_{1},e_{4},e_{5})$, $%
(e_{0},e_{1},e_{7},e_{6})$, $(e_{0},e_{2},e_{4},e_{6})$, $%
(e_{0},e_{2},e_{5},e_{7})$, $(e_{0},e_{3},e_{4},e_{7})$ and $%
(e_{0},e_{3},e_{6},e_{5})$, constitutes a canonical basis for the
quaternions in one-to-one correspondence with $(\ell ,\imath ,\jmath ,k)$.
Hence the octonions have one real unit, $7$ imaginary units and the
multiplication of two octonions is noncommutative. Similarly as for
quaternions and coquaternions we can view an octonion $n_{a}\in \mathbb{O}$
as a complex number 
\begin{equation}
n_{a}=a_{1}\ell +o\mathcal{O\in \mathbb{C}}(o)  \label{pctn}
\end{equation}%
with real part $a_{1}$, imaginary part $\mathcal{O}$ and newly defined
imaginary unit, $o^{2}=-1$, 
\begin{equation}
o:=\frac{1}{\mathcal{O}}\dsum\nolimits_{i=1}^{7}a_{1}e_{1}\qquad \text{where
\ }\mathcal{O=}\sqrt{\dsum\nolimits_{i=1}^{7}a_{1}e_{1}}.
\end{equation}

In order to obtain a $\mathcal{PT}_{o}$-symmetry we require a $\mathcal{PT}%
_{e_{1}e_{2}e_{3}e_{4}e_{5}e_{6}e_{7}}$-symmetry in the canonical basis.

\subsection{The octonionic Korteweg-de Vries equation}

Taking now an octonionic field to be of the form $%
u(x,t)=p(x,t)e_{0}+q(x,t)e_{1}+r(x,t)e_{2}+s(x,t)e_{3}+t(x,t)e_{4}+v(x,t)e_{5}+w(x,t)e_{6}+z(x,t)e_{7}\in 
\mathbb{O}$ the symmetric octonionic KdV equation, in this form of (\ref%
{KdVco}) becomes a set of eight coupled equations 
\begin{equation}
\quad 
\begin{array}{r}
p_{t}+6pp_{x}-6qq_{x}-6rr_{x}-6ss_{x}-6tt_{x}-6vv_{x}-6ww_{x}-6zz_{x}+p_{xxx}=0,
\\ 
\chi _{t}+6\chi p_{x}+6p\chi _{x}+\chi _{xxx}=0,%
\end{array}
\label{KdVoct}
\end{equation}%
with $\chi =q,r,s,t,v,w,z$. Setting any of four variables for $\chi $ to
zero reduces (\ref{KdVoct}) to the coupled set of equations corresponding to
the symmetric quaternionic KdV equation (\ref{qKdV}) with constraints (\ref%
{conq}).

\subsection{$\mathcal{PT}_{e_{1}e_{2}e_{3}e_{4}e_{5}e_{6}e_{7}}$-symmetric
N-soliton solutions}

Using the representation (\ref{pctn}) we proceed as in subsection \ref%
{comsol} and consider the shifted solution (\ref{us}) in the complex space $%
\mathbb{C}(o)$ 
\begin{eqnarray}
u_{a_{1}\ell +o\mathcal{O},\alpha }(x,t) &=&p_{a_{1},\mathcal{O};\alpha
}(x,t)-oq_{a_{1},\mathcal{O};\alpha }(x,t)  \label{sooct} \\
&=&p_{a_{1},\mathcal{O};\alpha }(x,t)\ell -\frac{1}{\mathcal{O}}q_{a_{1},%
\mathcal{O};\alpha }(x,t)\dsum\nolimits_{i=1}^{7}a_{1}e_{1}
\end{eqnarray}%
that solves the\ octonionic KdV equation (\ref{KdVoct}). The solution in (%
\ref{sooct}) is $\mathcal{PT}_{e_{1}e_{2}e_{3}e_{4}e_{5}e_{6}e_{7}}$%
-symmetric. Once more, multi-soliton solutions can be constructed in analogy
to the complex case $\mathbb{C}(\imath )$ treated in \cite{CenFring} by
treating all functions in $\mathbb{C}(o)$ as explained in more detail at the
end of section 4.

\section{Conclusions}

We have shown that the bicomplex, quaternionic, coquaternionic and
octonionic versions of the KdV equation admit multi-soliton solutions. Using
the standard folklore we assume that the existence of such type of solutions
indicates integrability of these equations, which we did not formally prove.
The bicomplex versions, local and nonlocal, display a particularly rich
structure with the two types of solutions found to exhibit very different
types of qualitative behaviour. Especially interesting is the solution in
the idempotent representation that decomposes a $N$-soliton into a $2N$%
-solitonic structure. Each one-soliton constituent of the $N$-soliton has
two contributions that even involve two independent speed parameters. Unlike
as for the real and complex solitons, where the degeneracy poses a
nontrivial technical problem \cite{CorreaFring,CCFsineG}, here these
parameters can be trivially set to be equal.

For all noncommuative versions of the KdV equation, i.e. quaternionic,
coquaternionic and octonionic, we found multi-soliton solutions based on
complex representation in which the imaginary unit is built from specific
combinations of the imaginary and hyperbolic units. Interestingly in all
cases we observe that the $\mathcal{PT}$-symmetry needed to ensure that the
newly defined imaginary unit can also be used as a $\mathcal{PT}$-symmetry
imposes constraints that are equivalent to the constraints needed to obtain
the symmetric KdV equation from the nonsymmetric one.

Naturally it would be interesting to extend the analysis presented here to
other types of nonlinear integrable systems. A more challenging extension is
to multi-complexify also the variables $x$ and $t$ which then also impacts
on the definition of the derivatives with respect to these variables.

\bigskip 

\noindent \textbf{Acknowledgments:} JC is supported by a City,
University of London Research Fellowship.

\newif\ifabfull\abfulltrue


\end{document}